\documentclass[5p]{elsarticle}

\usepackage{graphicx}
\usepackage{wrapfig}
\usepackage{enumerate}
\usepackage{textcomp}	
\usepackage{float}
\usepackage{dblfloatfix}
\usepackage{breakcites}

\usepackage{amssymb}

\usepackage{url}

\usepackage[hidelinks]{hyperref}

\begin{document}

\journal{arxiv.org}
\begin{frontmatter}

\title{Quartierstrom - Implementation of a real world prosumer centric local energy market in Walenstadt, Switzerland}

\author[mymainaddress]{L. Ableitner}
\author[mysecondaryaddress]{A. Meeuw}
\author[mymainaddress]{S. Schopfer}
\author[mymainaddress]{V. Tiefenbeck}
\author[mysecondaryaddress]{F. Wortmann}
\author[mymainaddress]{A. W\"orner}

\fntext[myfootnote]{Swiss legislation does not currently support such novel location-grid pricing schemes. To test this novel bottom-up tariff model in the absence of regulatory support, the grid costs from higher grid levels are covered with the research-project budget.}


\address[mymainaddress]{ETH Zurich, Chair of Information Management, Weinbergstrasee 56/58, CH-8006 Zurich}
\address[mysecondaryaddress]{University of St.Gallen, Institute of Technology Management, Dufourstrasse 40a, CH-9000 St.Gallen}

\begin{abstract}
Prosumers in many regions are facing reduced feed-in tariffs and currently have no possibility to influence the level of remuneration for the locally produced solar energy. Peer-to-peer communities may offer an alternative to the feed-in tariff model by enabling prosumers to directly sell their solar energy to local consumers (possibly at a rate that is beneficial for both consumer and prosumer). The Quartierstrom project investigates a transactional energy system that manages the exchange and remuneration of electricity between consumers, prosumers and the local electric grid provider in the absence of intermediaries. This whitepaper describes the prototypical real-world system being implemented in the town of Walenstadt, Switzerland, with 37 participating households. The community members of this pilot project pay a reduced tariff for grid usage if the electricity produced by a prosumer is sold to another community member, which is located on the same voltage or grid level downstream a substation\footnote{}. Such a tariff structure incentivizes local balancing, i.e. locally produced energy can be consumed locally whenever possible to avoid costs from higher grid levels. The blockchain is a novel technology suitable to log the produced and consumed units of energy within a community, making it possible to implement market places. In those marketplaces, both prosumers and consumers can indicate a price at which they are willing to sell / buy locally produced solar energy without the intermediation of a utility. The key goals of this project are the assessment of A) the technical, economical and ecological feasibility of a blockchain-based community energy system regarding local utilization of solar energy, grid quality and energy efficiency and B) resulting dynamics regarding local market prices and user acceptance. 
\end{abstract}


\end{frontmatter}



\section{Introduction}

\subsection{Solar power integration in electrical grids using peer-to-peer markets}

In most countries the electric power system relies heavily on fossil and nuclear fuels: About 61.4\% of the globally produced electricity was generated from gas and coal fired power plants in 2016. Nuclear power stations accounted for 10.4\% and hydro power, as main renewable source, accounted for 16.7\% of the annual electricity generation \cite{ieawebsite}. Many countries have put forward ambitious targets to reduce the dependency on fossil fuels, not just for electricity generation but for the entire energy system (including mobility and heating sectors). 
With current technology, the mobility and heating sector can be best decarbonized by electrification, which will further increase the electricity demand and requires the integration of more renewable generators. 

Photovoltaic (PV) systems are seen as a cornerstone of these ambitious decarbonization plans. Spurred by a rapid price decline with prices for residential PV systems falling by over 80\% from 2008 to 2016 in most competitive markets \cite{ECpvStatusReport2017}, PV systems have recently experienced a considerable increase in market diffusion in many countries, representing almost half of newly installed renewable power capacity in 2016 \cite{ECpvStatusReport2017}. 

Replacing large capacities with solar power will necessarily disrupt the power sector because solar power systems must be deployed in a fairly decentralized way, ranging from installation of larger utility-scale systems to small-scale, residential roof-top systems. Therefore, households installing roof-top solar systems turn from pure consumers into prosumers and cover parts of their electricity demand with solar energy. Decentralizing parts of the previously centralized generation capacity also implies a shift from large (in many cases public) investors to smaller private investors. For an extensive diffusion of decentral generation capacity it is therefore imperative that the profitability of decentralized solar systems can be guaranteed so that small private investors can contribute to replacing large, centralized power systems \cite{SchopferThesis}. 

Already today, the generation costs of solar energy are in many countries below the costs of  grid supplied electricity. However, prosumers generally cannot self-consume their entire solar generation. Today, they have no option but to feed their excess electricity into the public grid \cite{Luthander2015}. Therefore, the profitability of a PV system depends also on the remuneration (or feed-in tariff) for solar energy fed into the grid. 

Today, remuneration rates for solar energy are subsidized or regulated by the local grid operator in most countries\cite{schopfer2018economic}. Prosumers are therefore price takers and have no say in determining the level of the remuneration rate. In many regions, remuneration rates declined in the past years or are likely to decline in the future \cite{karneyeva2017solar}. This creates uncertainty regarding the profitability of solar systems and makes investments into PV systems less attractive due to unclear amortization rates in the future.
Peer-to-peer (P2P) energy markets have been proposed as an alternative approach that allows consumers and prosumers to directly trade surplus generated electricity within a microgrid and to individually set the price at which they are willing to trade the locally generated electricity. This shifts their role from passive price takers to an empowered role as active market participants who can influence the purchase and sales price of the electricity they trade with the other members of the microgrid.

\subsection{Network tariffs for the prosumer era}

Grid operators split up their operating and capital costs by voltage levels \cite{Johnston2017,Haro2017}. The highest network level connects large central power plants, which transport energy down to the low-voltage level networks, where households are located and connected to the power grid, as Figure 1 illustrates. Traditionally, electricity is being produced at large power plants connected to higher voltage levels and eventually distributed to customers at lower voltage levels. Consequently, the grid tariff comprises the costs arising at several voltage levels, following a top-down pricing scheme. Therefore, households are billed for both the consumed energy and the usage of the entire grid infrastructure. 

\begin{figure}
\centering
\includegraphics[width=100mm,scale=0.6]{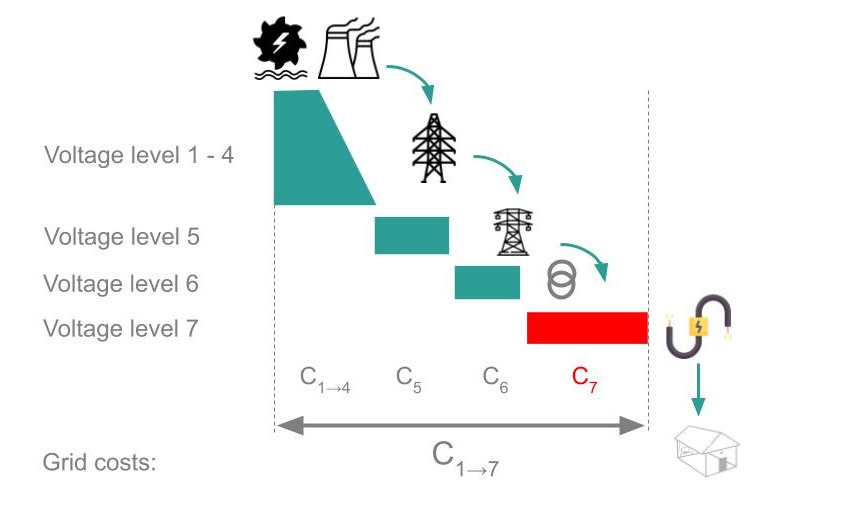}
\caption{Grid cost allocation implemented in Switzerland. The costs accumulate from the voltage levels 1 to 7 with costs $C_{1\rightarrow 7}$.}
\label{fig:grid_cost_allocation}
\end{figure}

With the ongoing decentralization, electrical energy is increasingly generated on a neighborhood or district scale (voltage level 7 in Figure \ref{fig:grid_cost_allocation}). The traditional top-down cost allocation does not incentivize local consumption of solar energy beyond self-consumption within a household. This implies that a consumer is billed the same network fee regardless of whether the energy has been generated from a solar system in the neighbourhood or from a distant power station. 

To create an incentive for the consumption of locally generated electricity, locational grid pricing could be implemented that takes the shorter trading distances and the non-use of higher voltage levels into account. Consequently, if a unit of generated solar energy is consumed by a household within the same microgrid (i.e. below the same substation), only the grid costs of voltage level 7 $C_7$ apply. If no trades occur within the community, the full network costs $C_{1\rightarrow 7}$ apply. Such tariff schemes allocate costs on bottom-up fashion rather than a top-down fashion. They automatically incentivize local balancing by expanding the optimization of self-consumption from the household level to the community/microgrid level.

\subsection{Towards prosumer-centric ecosystems}

For prosumers, the active market participation within a microgrid with bottom-up grid tariff provides an ecosystem that potentially allows profitable operation of solar systems in the absence of subsidized remuneration rates. However, the concept of a peer-to-peer system with bottom-up grid tariff is quite novel and not possible under the current legislation in many regions. Hence, there is limited experience with real-life pilot tests that analyze to what extent a decentralized market fulfills prosumer and consumer needs over traditional and regulated approaches with fixed remuneration rates or feed-in tariffs. 

\section{The Quartierstrom project: scope and test site}

The ``Quartierstrom'' (German for district power) is supported by the Swiss Federal Office of Energy (SFOE) within the framework of its pilot, demonstration and flagships programme. It is one of the first projects that implements and evaluates the feasibility of a decentral local P2P energy market in a real-world pilot test with 37 households located in the town of Walenstadt, Switzerland. Together with partners from industry and academia \cite{qsWebsite}, the project assesses: 
 \begin{enumerate}[A.]
 \item The \textbf{technical feasibility} of a blockchain-managed community energy system and its impact in terms of local utilization of solar energy, grid quality and energy efficiency
\item Suitable \textbf{market mechanisms} and resulting market prices for local electricity
\item The design of an appropriate \textbf{user interface}, user engagement over time and the overall acceptance of the system by the users.
 \end{enumerate}

The local community for the pilot test consists of 37 participating households that are located downstream of a substation (grid level 7). The community consists of 27 prosumers with a total solar PV capacity of 280 kWp with a total storage capacity of 80 kWh in form of lithium ion batteries. As part of the project implementation, each of the households has been equipped with prototypical smart-meters that can measure currents, voltages and frequencies on each of the phases and that come with an integrated single board computer (SBC). The SBCs act as distributed nodes for the blockchain system that is introduced in chapter 4. Each household is equipped with a maximum of three smart meters to separately measure net consumption, production and battery power.

The remainder of this whitepaper describes pilot set up and the three main components of the field test:
 \begin{enumerate}[1.]
 \item Market design and mechanism
 \item Blockchain design and technical implementation
 \item User interface design and implementation
 \end{enumerate}

\section{Market design and mechanism}

In order to create a successful peer-to-peer exchange for electricity in line with the overall objective of promoting sustainability, there should be incentives for local consumption of locally generated electricity. Consequently, prices should reflect the availability of local electricity. The market design describes the way in which prices are determined and local electricity is distributed within the community and should be defined with the aim to meet these overall project goals. 

The importance of an appropriate market design for electricity distribution is illustrated by the infamous example of the breakdown of the California electricity market in 2001, which caused unforeseen price jumps and even led to outages \cite{borenstein1995market}. In general, there are two alternatives for determining prices and the allocation of electricity: a) running an algorithm which calculates prices relative to the availability of local resources and distributes the available energy randomly to the community members; b) running an auction mechanism that allows the participants to state price limits for which they are willing to buy or sell electricity with-in the local grid. 
Most existing electricity markets are governed by auction mechanisms \cite{rosen2013auction}. That means that participants express their preferences in bids which contain a price and a commodity or quantity of commodities they wish to buy. All bids are collected in an order book and at distinct times, these orders are matched to form trades between the participants according to specific auction rules. These auction rules have a strategic impact on the incentives market participants face when formulating their bids in order to maximize the expected benefits. There are numerous research studies on auction mechanisms and their implications for economic efficiency and price development.

However, those existing markets are wholesale markets, in which professional firms interact instead of private households as is the case in the Quartierstrom peer-to-peer market. It is a novel situation that the households, which are currently merely price-takers in a retail market, change their roles to active prosumers and consumers which influence electricity sourcing themselves. 
Based on economic theory it is possible to estimate the evolution of prices and consumption within the community. What is unique about this project is that the implementation of a market on a distributed system in the field with real participants gives us the opportunity to test the individual's participation and true willingness to pay in the field. This allows us to get an understanding of the participants' behavior and to find out whether a decentralized market can really govern itself efficiently in this domain. 

A review of existing studies and companies engaged in building in peer-to-peer electricity markets that we conducted leads us to the conclusion that except for few known exceptions, most projects are far away from implementing a system in the field. The blockchain platform as well as the market design or pricing mechanism are described in very fuzzy terms and actual user engagement/a real-world user interface is hardly ever mentioned. This is due to the fact that most projects are at a stage in which they only present concepts or develop first proofs of concept in the lab, but are not yet rolling out their solutions in the field with real participants.

\subsection{Mechanism Design}
A key decision in the market design was to create an auction in which all participants have the possibility to influence the prices for which they buy or sell electricity. While we do not necessarily expect all participants to adapt their prices frequently in the long run, we do believe this is a unique chance to elicit price preferences for local, renewable energy from individuals in a real setting. There exist some studies which have asked individuals about their willingness to pay for green or local electricity or about their intentions to invest in renewable energies, e.g. \citep{Ecker2018}. But to the best of our knowledge, this is the first study which allows individuals to actually influence the true prices they will pay. \\
Based on the experiments and studies resulting from the literature research, we have identified a double auction with discriminative pricing as the most suitable market mechanism for the Quartierstrom market. For both, consumers and prosumers, the smart meters send bids containing the price limit determined by the individual household and the electricity demand or supply measured by the smart meters. An order book collects all bids during an interval of 15 minutes and orders them by price: Sell bids with a lower sell price are prioritized (see yellow curve in figure below), buy bids with a higher price (see purple curve). Discriminative pricing means that for each trade, the price is derived as the mean between the respective buyer's and seller's price bid (see turquoise dotted line). This auction is run iteratively every 15 minutes, which means that the market is cleared and trades are allocated in this frequency. 

\subsection{Implementation at the pilot site}
We have deployed this double auction which will run throughout the rest of the field test. The double auction has been implemented in TypeScript and has been integrated into the blockchain platform. The implementation integrates the local utility company, the WEW Walenstadt, in the market by assigning all excess bids which cannot be filled with local supply or demand to the utility provider at existing tariffs. All of the blockchain nodes run the auction algorithm every 15 minutes to clear the market. To enable users to set their desired limit prices for the auction, we have implemented a price slider in the user interface in the WebApp described in section \ref{user_interface}.

\section{Blockchain design and technical implementation}

\subsection{Blockchain infrastructure}

With the goal of a local peer-to-peer energy market in mind, we have built a system that is distributed and secured by the participants and beneficiaries of it.
We have chosen a blockchain-based approach to allow for mutual validation of the correctness of transactions, computation, and settlement of the system without requiring a central authority.
As Fig. \ref{fig:system_overview} shows, the system consists of three different types of participants.
The system must be able to scale to larger numbers of participants. The core of the platform is built by its validator nodes, which are represented by the prosumers and the utility company and directly hosted on the smart-meter with integrated single-board computer.
The user base is represented by the consumer nodes, which are clients of the prosumers and do not propose new blocks on their own.
Consensus over the current state of the system is negotiated between the validating nodes using the Tendermint consensus protocol \cite{tendermint}.
Tendermint allows for replicated state machines to be kept between an arbitrary number of validators.
Based on the tendermint protocol, self-sovereign and interoperable proof-of-stake systems can be constructed that do not require extensive electricity demand for block validation.

Every participant runs either a full or a light node to connect to the underlying blockchain, as well as the smart meter and agent modules. Both full and light node are hosted directly on a smart-meter with integrated single board computer. In order to offer services to the end-users, such as comprehensible information about consumption and production via external applications, a block explorer is available to query the system for addresses and transactions.

\begin{figure}[htb]
  \centering
  \includegraphics[width=95mm,scale=0.5]{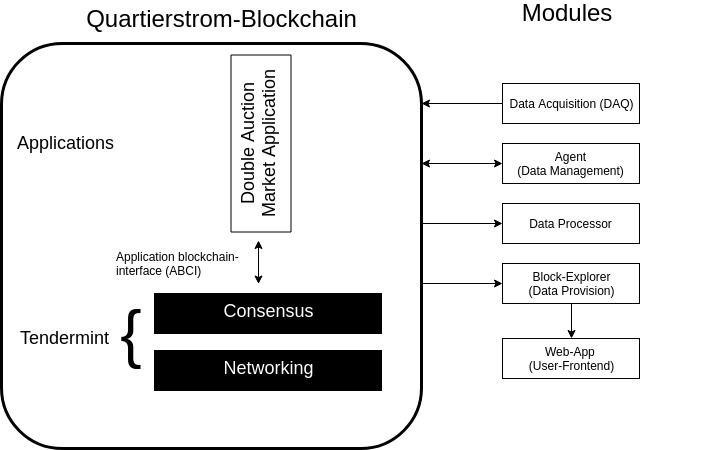}
  \caption{Platform and blockchain application structure}
  \label{fig:platform}
\end{figure}

\subsection{Funcional modules}
Figure \ref{fig:platform} gives an overview of the platform and the fundamental modules and building blocks. We use Tendermint for consensus and networking. Any application can be hosted on top of Tendermint. The first application being implemented is the peer-to-peer (P2P) market application along with the modules that interact with the blockchain.
The market application needs to be able to receive bid data from the participating nodes and execute functions, like clearing and settling, in a fixed interval. The functionality can be split up into four distinct parts of the system which are part of every participant of the system:
 
\begin{itemize}
\item Data Acquisition: Smart meter and read-out application
\item Data Management: Agent and client application to process acquired data, issue transactions and manage signatures
\item Data Processing: Full / light node for execution and validation of platform applications and subscription to updates
\item Data Provision: Making data available for applications and user interfaces
\end{itemize}

\begin{figure*}[b]
  \centering
  \includegraphics[width=160mm,scale=0.5]{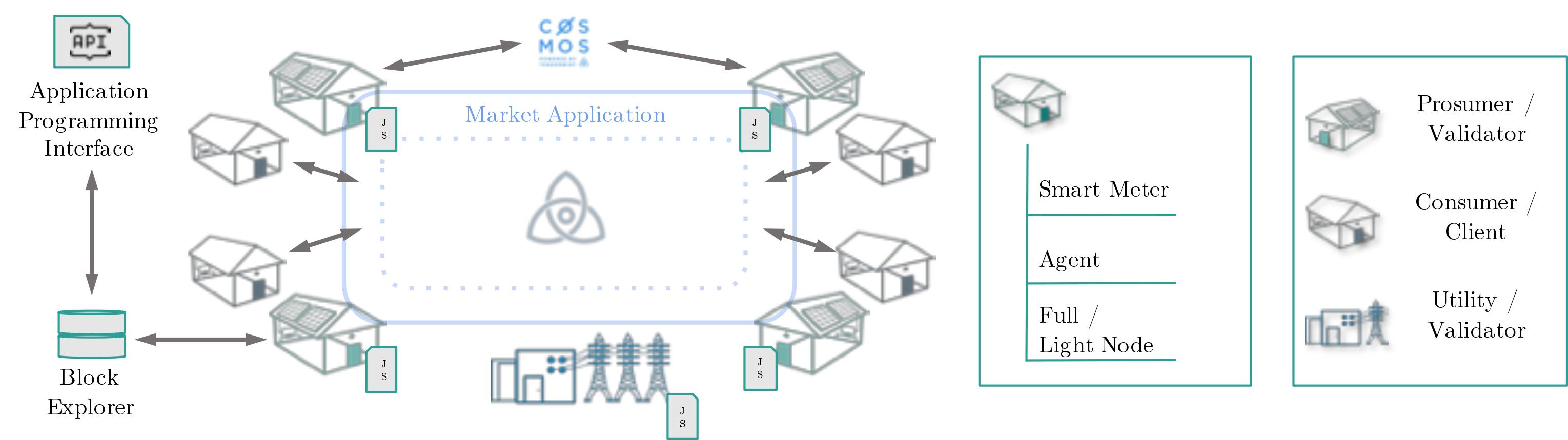}
  \caption{System overview. This figure shows the three categories of network participants: prosumers, consumers, and the utility company. The three modules for data acquisition (smart meter), data management (agent), and data processing (full/light) node are implemented on every smart meter of every participant.}
  \label{fig:system_overview}
\end{figure*}

The availability of the platform and of its applications is vital to its operation: A centralized platform architecture requires that a server is always available (i.e. in operation), while our decentralized platform requires that 2/3 of non-malicious nodes are available (Byzantine fault tolerance property of Tendermint \cite{tendermint}) to perform validation for the platform to operate.
The Byzantine fault tolerance property of the platform increases the resiliency of the system and allows nodes to continue safe operation as long as no more than 2/3 of the nodes are offline or malicious.
Such fault tolerant behavior is especially useful for nodes that control flexible assets, which in turn may be used in the future curtail solar energy or demand peaks to reduce loads and over voltages in the distribution grid.

\paragraph{Data Acquisition} Even though data acquisition is separate from a blockchain-based functionality, it is part of the chain of trust, as measurement data is the basis for settlements in the market application. 
The smart meters (Smartpis) are not only responsible for the data acquisition but host the P2P market application (i.e. the market application runs on the smart meter) as well as the \textit{Data Management} and \textit{Processing} functionalities of the platform.

\paragraph{Data Management}
The management of actions, such as sending out buy or sell orders or updating price preferences based on the acquired data, is handled on each of the end-users\textquotesingle s devices.
This functionality is separate from being a validator or just a subscriber to the system.
The agent module performs the coordination of actions on the end-user\textquotesingle s side of the system.
While synchronizing with the blockchain and its regularly published blocks, this module contains information about the user\textquotesingle s preferences, such as sell and buy prices, and follows a strategy according to its collected information.
The strategy contains descriptions of the agent's objectives and issues buy or sell orders based on the current consumption and production.
The modules in Figure \ref{fig:platform} contain the schemes for data types as well as methods for composing and signing a transaction.
The transaction scheme includes the sender's public key as well the sender's address, similar to values typically found in most blockchain transactions.
The module provides methods for generating new private keys, deriving public keys from existing private keys, as well as generating addresses from them, and uses the same Elliptic Curve Cryptography (ECC) curve as in Bitcoin (secp256k1).
By using the SHA256 hash of a user's public key as an address, we enable our transactions to have a unique identifier while preserving the privacy of a user by not being able to be traced.

\begin{figure*}[b]
  \centering
  \includegraphics[width=160mm,scale=0.5]{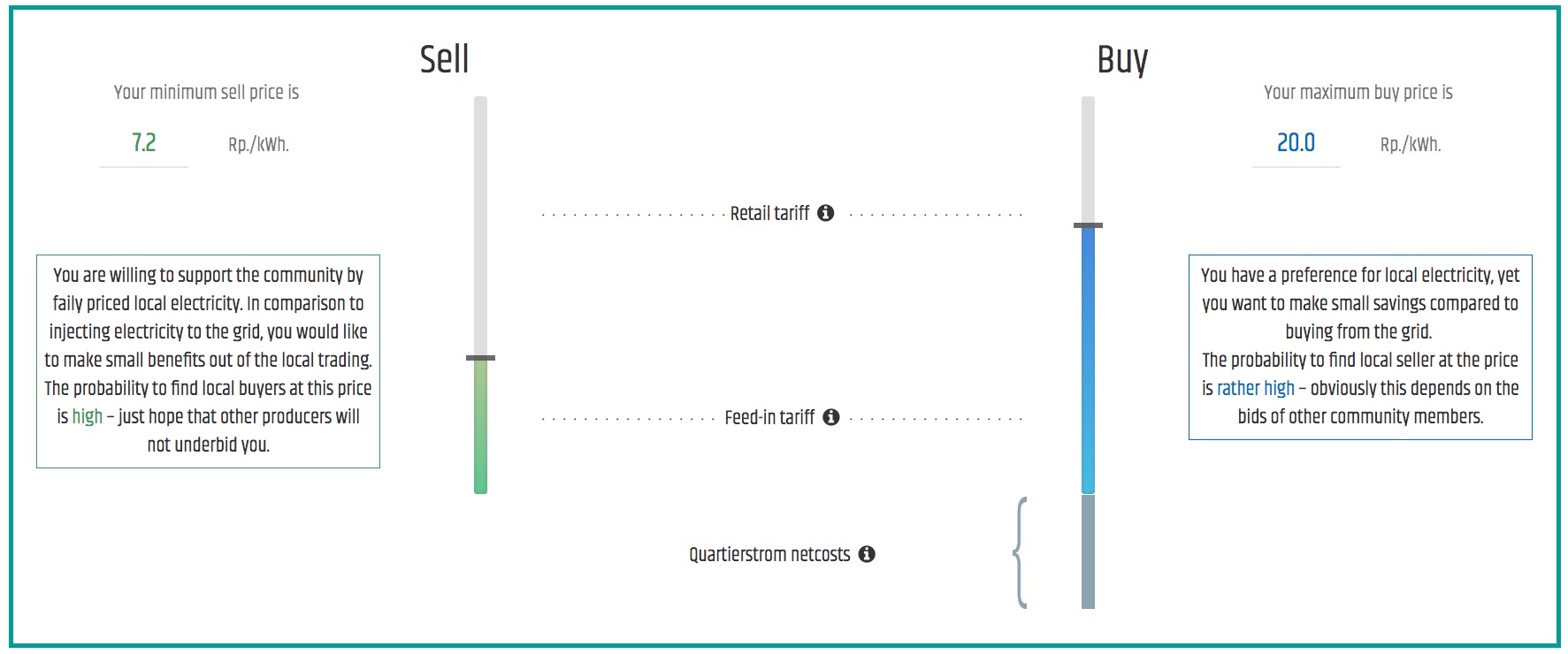}
  \caption{Price slider functionality in the web-app}
\end{figure*}

\paragraph{Data Processing}
The issuance and signature authority within the platform's BFT based a proof-of-stake consensus mechanism is distributed to the utility company and prosumer participants.
The Tendermint consensus protocol allows for high adaptability due to its Application Blockchain Interface (ABCI) universally available to any programming language.
In addition, Tendermint offers a high amount of flexibility and customizability in order to adjust to particular application requirements such as: the reduction of communication, creation of empty blocks, and time delay between blocks.
The initial decision to trust prosumers in the systems is based on the rationale that that these participants have already made an investment in the system (in the form of a PV system).
While the current distribution of voting power is equal for every active validating node, future enhancements of the platform could include an active staking mechanism to incorporate solar investment size while maintaining a valid equilibrium between the nodes.
Blocks are created periodically.
By checking all blocks via the \textit{BlockHandler}, a self-triggering mechanism within the platform allows for automated actions as required by the market application.
Incoming transactions are handled by the \textit{TransactionHandler} according to their receiver and payload.
When verifying the validity of a transaction, the node deserializes its contents, derives the address from its contained public key, and compares it with the given address.
If the addresses match, the client hashes the contents of the transaction and verifies the transaction signature against the sender's public key.
According to the receiver address and the data contained in the payload field of the transaction, the transaction handler forwards the information to the respective handlers of the application modules.
In case of the market application, the arrival of a transaction containing a sell- or buy-order triggers the addition of a new bid.

\paragraph{Data Provision} In order to offer services to the end-users, such as comprehensible information about consumption and production via external applications (such as the user-interface), a block explorer is available to provide queryable access to the blockchain data for addresses and transactions.

\section{User interface}
\label{user_interface}

\subsection{Users as key players in P2P energy markets} 
The end users of a P2P energy market are the key players in the kind of auction model described above. By signing up for participation in a local energy market, all participants can contribute to the sustainable energy production and its local consumption within the community (by buying and selling locally). With an active market participation by setting limits for buy prices or sell prices, users make use of their right of self-determination and participate in the pricing of local electricity. The latter requires a user interface (UI) that makes trading with the neighbors easy and attractive. 
While P2P electricity trading is currently widely discussed in research and practice, actual prototypes with user interfaces (UIs) of P2P energy are still very scarce. The start-up LO3 recently released their interface for the Brooklyn Microgrid \cite{lo3}. It primarily focuses on identifying roofs in the neighborhood to further expand the community, providing insights into energy data, and on defining the maximum amount users are willing to spend for electricity. Some other start-up projects in the field claim to also develop UIs for P2P electricity trading, but so far have not released them to the public. A recent project in Human-Computer-Interaction research (HCI) explores the potential of energy transaction data; in addition to providing feedback to consumers on their electricity consumption, the projects visualizes the origin of the electricity in a spatial representation \cite{meeuw2018lokalpower}. However, the UI was not actually used in the field and potential users emphasized the importance of privacy in a location-based approach.  

\subsection{Development and presentation of the UI} 
In the development of the Quartierstrom UI, we started from the user perspective, collecting specific user needs from which we derived design implications. The user needs were collected in so-called focus groups, a form of group interview to which we invited prosumers and consumers separately. Focus groups offer in-depth qualitative insights into the user perspective and benefit from interactions between participants. Especially the latter is important in the energy context: in contrast to surveys which are usually answered in isolation, the surrounding atmosphere of focus groups might in particular encourage participants who are less knowledgeable or who are not particularly interested in energy topics to start forming an opinion and expressing perceived benefits and barriers. After all, energy is a challenging topic for user engagement: it is invisible, intangible, and a commodity taken for granted; consequently, the topic is not so relevant to most consumers. 

From the workshop results, we derived the four key functionalities and design principles for user interfaces of P2P energy markets as listed in Table \ref{tab:DSF}

\begin{table}[h]
\centering
\begin{tabular}{p{1cm}|p{3.2cm} | p{3.2cm}|}
\hline
\textbf{Nr.} & \textbf{\textbf{Domain-specific functionalities (DSF)~}}  & \textbf{\textbf{General design principles \quad}}  \\
\hline
DSF1 & Empower users to set a P2P price                               &  Highlight value proposition of P2P electricity    \\\hline
DSF2 & Provide transparency in electricity consumption and production &  Design for ease-of-use                            \\\hline
DSF3 & Provide financial transparency                                 &  Tailor content to information needs               \\\hline
 DSF4 &
 Enable community relatedness                                   & 
Have a strategy for continuous usage     
\\ \hline        
\end{tabular}
\caption{Listing of domain specific functionalities (DSF) and general design principles\label{tab:DSF}}
\end{table}

Based on these core functionalities and design principles, we developed wireframes, reviewed them in our team and discussed them with industry experts. As interface type, we chose a webapp which is easily accessible through all standard web browsers. The webapp itself is based on \textquotesingle s Flask library, a bootstrap dashboard template that we adapted heavily for our purposes and customized HTML and Javascript code.  
In the following we present a sneak preview of the user interface:

\begin{figure}[h]
\centering
\includegraphics[width=85mm,scale=0.5]{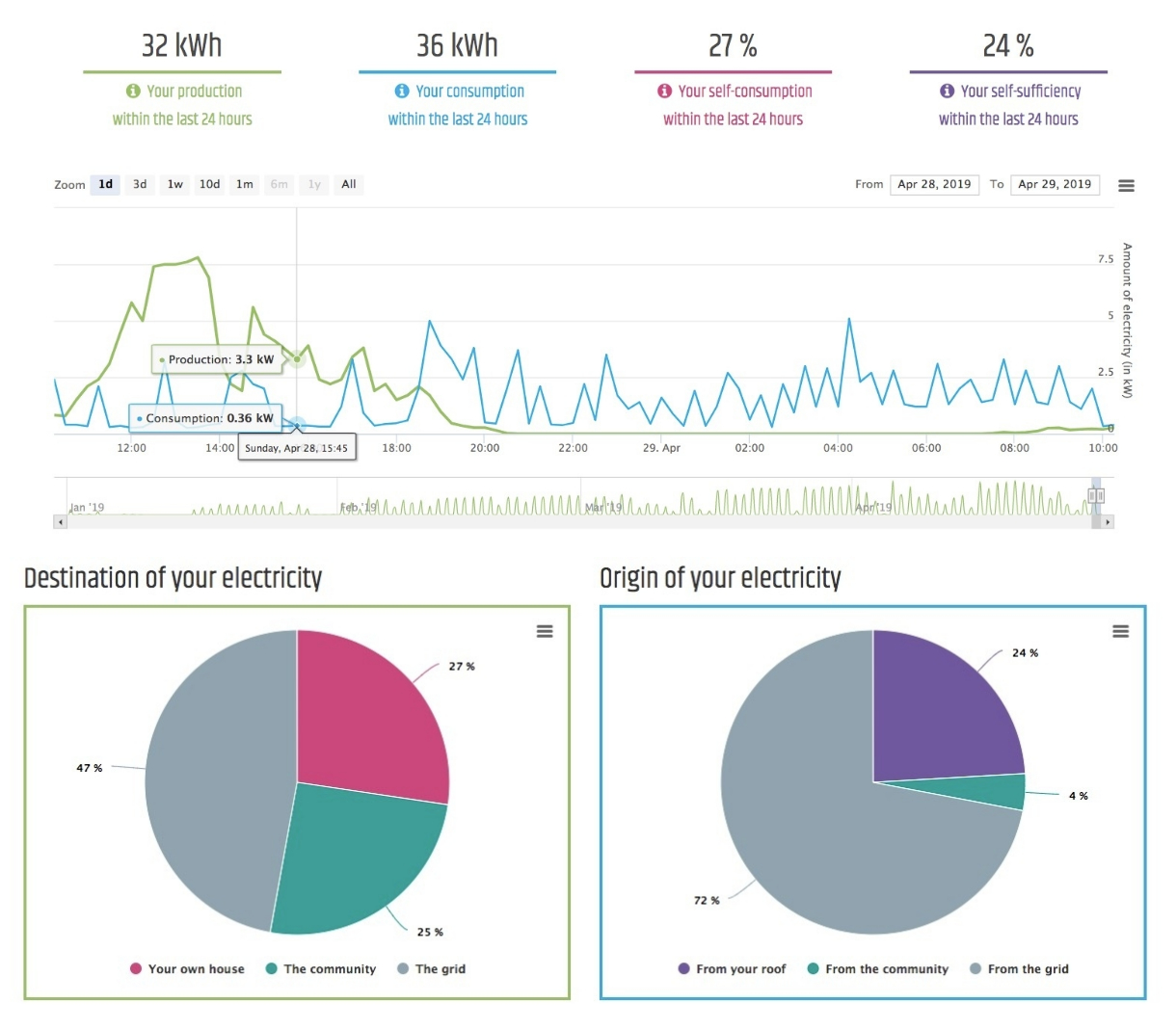}
\caption{Personal energy data}
\end{figure}

\bigskip

\textbf{The price sliders} (DSF1) 
Quartierstrom grants its users a right of self-determination and enables them to participate in the pricing of local electricity. Prosumers set a minimum limit for selling electricity to their neighbors (green slider) and indicate Pythontheir willingness to buy local electricity from their neighbors (blue slider).  Consumers set only their maximum purchase price. Information boxes next to the sliders inform users about the probability of selling, resp. buying electricity locally based on their price limits. The limits are active until the users changes them.  
Every 15 minutes, the agent transforms the user's limits in combination with the electricity consumption measured by the smart meters into a bid on the local energy market. Buyer and seller are matched, payment is settled and the transaction is broadcast to the network. 
\iffalse
\begin{tabular}{ll}
\raisebox{-.5\height}
{\textbf{Insights into personal energy data} (DSF2) 
Users are provided with real-time and historic information on their electricity production (green) and demand (blue). A time series graph is accompanied by summary key performance indices (KPIs) on production, demand, self-consumption ratio and self-sufficiency ratio. Both, the graph and the KPIs are interactive on mouse-overs and update with the adjustment of the timeframe. Users can choose timeframes at given intervals (buttons on the left side), with a calendar feature (right) or a zoom band (below). & & \\
\end{tabular}

\begin{thebibliography}{10}
\expandafter\ifx\csname url\endcsname\relax
  \def\url#1{\texttt{#1}}\fi
\expandafter\ifx\csname urlprefix\endcsname\relax\def\urlprefix{URL }\fi
\expandafter\ifx\csname href\endcsname\relax
  \def\href#1#2{#2} \def\path#1{#1}\fi

\bibitem{ieawebsite}
IEA, \href{https://www.iea.org/statistics/electricity/}{Electricity
  statistics}.
\newline\urlprefix\url{https://www.iea.org/statistics/electricity/}

\bibitem{ECpvStatusReport2017}
{European Commission},
  \href{publications.jrc.ec.europa.eu/repository/bitstream/JRC103426/ldna28159enn.pdf}{{PV
  Status Report 2016}}, Tech. rep. (2016).
\newline\urlprefix\url{publications.jrc.ec.europa.eu/repository/bitstream/JRC103426/ldna28159enn.pdf}

\bibitem{SchopferThesis}
S.~Schopfer,
  \href{https://www.research-collection.ethz.ch/handle/20.500.11850/341266}{Assessment
  of the consumer-prosumer transition and peer-to-peer energy networks}, Ph.D.
  thesis (2019).
\newline\urlprefix\url{https://www.research-collection.ethz.ch/handle/20.500.11850/341266}

\bibitem{Luthander2015}
R.~Luthander, J.~Wid{\'{e}}n, D.~Nilsson, J.~Palm, {Photovoltaic
  self-consumption in buildings: A review}, Applied Energy 142 (2015) 80--94.
\newblock \href {http://dx.doi.org/10.1016/j.apenergy.2014.12.028}
  {\path{doi:10.1016/j.apenergy.2014.12.028}}.

\bibitem{schopfer2018economic}
S.~Schopfer, V.~Tiefenbeck, T.~Staake, Economic assessment of photovoltaic
  battery systems based on household load profiles, Applied energy 223 (2018)
  229--248.

\bibitem{karneyeva2017solar}
Y.~Karneyeva, R.~W{\"u}stenhagen, Solar feed-in tariffs in a post-grid parity
  world: The role of risk, investor diversity and business models, Energy
  Policy 106 (2017) 445--456.

\bibitem{Johnston2017}
J.~Johnston,
  \href{http://linkinghub.elsevier.com/retrieve/pii/B9780128117583000164}{{Peer-to-Peer
  Energy Matching: Transparency, Choice, and Locational Grid Pricing}}, in:
  Innovation and Disruption at the Grid's Edge, Elsevier, 2017, pp. 319--330.
\newblock \href {http://dx.doi.org/10.1016/B978-0-12-811758-3.00016-4}
  {\path{doi:10.1016/B978-0-12-811758-3.00016-4}}.
\newline\urlprefix\url{http://linkinghub.elsevier.com/retrieve/pii/B9780128117583000164}

\bibitem{Haro2017}
S.~Haro, V.~Aragon{\'{e}}s, M.~Mart{\'{i}}nez, E.~Moreda, A.~Morata,
  E.~Arb{\'{o}}s, J.~Barqu{\'{i}}n,
  \href{http://linkinghub.elsevier.com/retrieve/pii/B9780128117583000127}{{Toward
  Dynamic Network Tariffs: A Proposal for Spain}}, in: Innovation and
  Disruption at the Grid's Edge, Elsevier, 2017, pp. 221--239.
\newblock \href {http://dx.doi.org/10.1016/B978-0-12-811758-3.00012-7}
  {\path{doi:10.1016/B978-0-12-811758-3.00012-7}}.
\newline\urlprefix\url{http://linkinghub.elsevier.com/retrieve/pii/B9780128117583000127}

\bibitem{qsWebsite}
Quartierstrom-consortium.
\newblock \href{http://www.quartier-strom.ch}{Quartierstrom website} [online]
  (2019).

\bibitem{borenstein1995market}
S.~Borenstein, J.~Bushnell, E.~Kahn, S.~Stoft, Market power in california
  electricity markets, Utilities Policy 5~(3-4) (1995) 219--236.

\bibitem{rosen2013auction}
C.~Rosen, R.~Madlener, An auction design for local reserve energy markets,
  Decision Support Systems 56 (2013) 168--179.

\bibitem{Ecker2018}
F.~Ecker, H.~Spada, U.~J. Hahnel,
  \href{https://doi.org/10.1016/j.enpol.2018.07.028}{{Independence without
  control: Autarky outperforms autonomy benefits in the adoption of private
  energy storage systems}}, Energy Policy 122~(July) (2018) 214--228.
\newblock \href {http://dx.doi.org/10.1016/j.enpol.2018.07.028}
  {\path{doi:10.1016/j.enpol.2018.07.028}}.
\newline\urlprefix\url{https://doi.org/10.1016/j.enpol.2018.07.028}

\bibitem{tendermint}
Tendermint.
\newblock \href{https://tendermint.com/docs/}{Tendermint documentation}
  [online] (2019).

\bibitem{lo3}
LO3.
\newblock \href{http://www.brooklynmicrogrid.com}{The brooklyn microgrid}
  [online] (2017).

\bibitem{meeuw2018lokalpower}
A.~Meeuw, S.~Schopfer, B.~Ryder, F.~Wortmann, Lokalpower: Enabling local energy
  markets with user-driven engagement, in: Extended Abstracts of the 2018 CHI
  Conference on Human Factors in Computing Systems, ACM, 2018, p. LBW613.

\end{thebibliography}
\end{document}